# Depth-resolved Nuclear Resonance Scattering under X-ray standing wave - an approach to study interface magnetism

Dileep Kumar

UGC-DAE Consortium for Scientific Research, Khandwa Road, Indore-452001, India

dkumar@csr.res.in

**Abstract**

The isotope selective grazing-incidence nuclear resonance scattering (GI-NRS) technique is demonstrated to be depth-resolved under x-ray standing wave (XSW) conditions to probe the magnetism of the two interfaces of the Fe layers (Fe-on-Tb and Tb-on-Fe interface) independently in Tb/Fe/Tb trilayer structures. Depth resolution was achieved by placing an ultra-thin layer of $^{57}$Fe at both interfaces (Tb/$^{57}$Fe/Fe/$^{57}$FeTb). Intentionally, both $^{57}$Fe layers were assumed to have different hyperfine fields and orientations. Based on theoretical simulations, it is demonstrated that the antinode regions of XSWs generated through the W/Si multilayer structure allow one to independently measure the Fe-on-Tb and Tb-on-Fe interface at different incident angles. These theoretical simulations of NRS patterns at different incident angles are found to correspond to $^{57}$Fe layers independently in the Tb/Fe/Tb trilayer. The present work shows the capability of combining XSW and GI-NRS to study buried magnetic interfaces in thin film structures.

## Introduction

Magnetic multilayers with perpendicular magnetic anisotropy (PMA) are strong candidates for application in perpendicular magnetic recording media. There has been a growing realization that the interface structure in such multilayers strongly affects their magnetic properties [1] For example, in the Fe/Tb multilayer, it has been inferred that the Fe-on-Tb interface is more diffused than Tb-on-Fe, and PMA originates mainly from the Fe-on-Tb interface [3]. Thus, the asymmetry in two types of interfaces viz; A-on-B and B-on-A may significantly affect PMA. More recently, systems like Co/Pt, Fe/Pt, and Fe/Pd have been found to possess strong PMA after appropriate post-deposition treatments. In both the classes of multilayers, the origin of PMA is not fully understood, although the structure of the interfacial region is known to play an important role in it [3].

To understand the role of interface structure in determining the magnetic properties in such systems, it is necessary to study i) both interfaces in identical conditions to correlate with existing PMA in the same sample ii) trilayer film structure to avoid the effect of averaging over large in-equivalent interfaces in case of multilayers structure. To the best of my knowledge, a direct magnetic measurement of both interfaces (A-on-B and B-on-A) independently in the same trilayer sample is not possible. Although various techniques have been used to get indirect information about the interfacial region, several unexpected results have been obtained in various systems [4–8]. The availability of high-brilliance synchrotron radiation has recently opened new avenues for the application of nuclear resonance scattering (NRS) methods in this field. Novel features of this technique, such as high sensitivity and isotope selectivity, could be used to differentiate very small changes in magnetic structure, such as the orientation and magnitude of the hyperfine field at both interfaces [9]. This can be done by placing an ultra-thin layer of nuclear isotope, such as $^{57}$Fe, at both interfaces. In addition to this, depth-resolved measurements under the x-ray standing wave (XSW) condition will allow us to resolve both interfaces independently in the same sample [10].

In the present theoretical simulation, depth-resolved grazing incident nuclear resonance scattering (GI-NRS) measurements will be used to determine the magnetic information of the interfacial region in the magnetic multilayer (Tb/Fe/Tb). The trilayer sample for the theoretical simulation will be prepared by taking an ultra-thin $^{57}$Fe layer at both interfaces of the Fe layer (Fe/$^{57}$Fe–on–Tb and Tb-on-$^{57}$Fe/Fe). Depth selectivity will be obtained by generating XSW using an underlying W/Si multilayer. These theoretical simulations of NRS patterns at different incident angles correspond to $^{57}$Fe layers independently in the Tb/Fe/Tb trilayer. The present work shows the capability of the combination of XSW and GI-NRS for the study of buried magnetic interfaces in thin film structures.

**Experimental**

The following sample structure is used for the present study.

*Substrate+(W/Si) multilayer/Si wedge/ **Tb/ $^{57}$Fe/ Fe(natural)/ $^{57}$Fe/Tb** + cap layer (Pt)*

When the x-rays are incident at an angle close to the Bragg angle of the W/Si multilayer, x-ray standing waves are generated in the multilayer, which extend beyond the mirror and into the trilayer [7]. As the angle of incidence varies across the width of the multilayer Bragg peak, the position of the antinode moves over the bilayer thickness. The antinode of the XSW overlaps with both resonant ($^{57}$Fe) layers one by one at two different incident angles [11]. In this way, independently, unambiguous information about both interfaces can be obtained by measuring the GI-NRS decay spectrum at different angles of incidence. It may be noted that a-Si wedge can also be used to vary the height of the $^{57}$Fe/Fe-on-Tb and Tb-on-$^{57}$Fe/Fe interfaces with respect to the standing wave pattern formed by the underlying W/Si mirror. In a recent study at ID-32 beamline, ESRF, France, we have demonstrated that XSW generated by a multilayer mirror can be used for a depth resolution of a fraction of a nanometre [10].

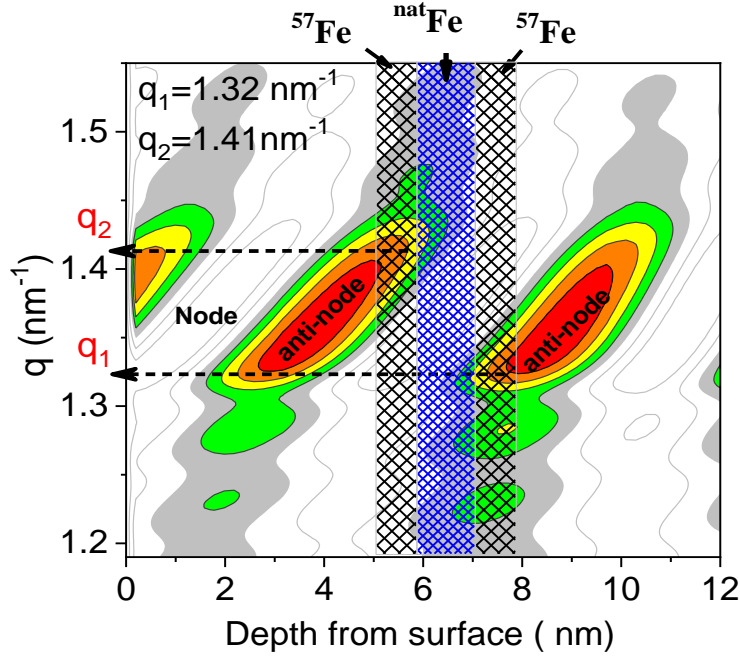

**Fig. 1:** Contour plot of the x-ray intensity distribution inside the tri-layer as a function of q

**Simulated results and discussions**

Feasibility of the experiment is demonstrated through the following simulation by taking the multilayer structure: *$Si_{sub}$/[W 2.0 nm/Si 3.0 nm]x 20/ $Si_{wedge}$/ Tb 2.0nm /$^{57}$Fe 0.8 nm/ Fe 1.2nm/$^{57}$Fe 0.8 nm/Tb 2.0 nm/Si 3 nm.* The contour plot in Fig. 1 shows the calculated x-ray intensity distribution inside the above-mentioned film structure as a function of q (red is the maximum field intensity at the antinode of the XSW) for a given value of the thickness (1.5 nm) of the $Si_{wedge}$ layer [1]. The position of the $^{57}$Fe layers and $^{nat}$Fe is shown as shaded bar. One may note that as the angle of incidence is varied across the width of the multilayer Bragg peak, $r^{th}$ and $r+1^{th}$ antinode crosses the top and bottom $^{57}$Fe layer respectively at the different incident angles ($q_1$ and $q_2$). At these q values NRS decay spectrum could be obtained preferentially from Tb-on-$^{57}$Fe and $^{57}$Fe-on-Tb interface independently. Since the GI- NRS technique is isotope sensitive, the weightage of the central $^{nat}$Fe layer relative to the two interfaces will be almost equal and negligible, which makes such experiments sensitive to the interface. It is also clear from the calculated GI-NFS decay spectrum simulated at these two q values by taking different magnitudes of hyperfine fields of $^{57}$Fe layers. It may be noted that the beat pattern with different periodicity in the theoretical simulation could be seen just by taking two different incident angles. These theoretical simulations show that the technique can resolve two interfaces reasonably in a single sample just by changing the incident angle.

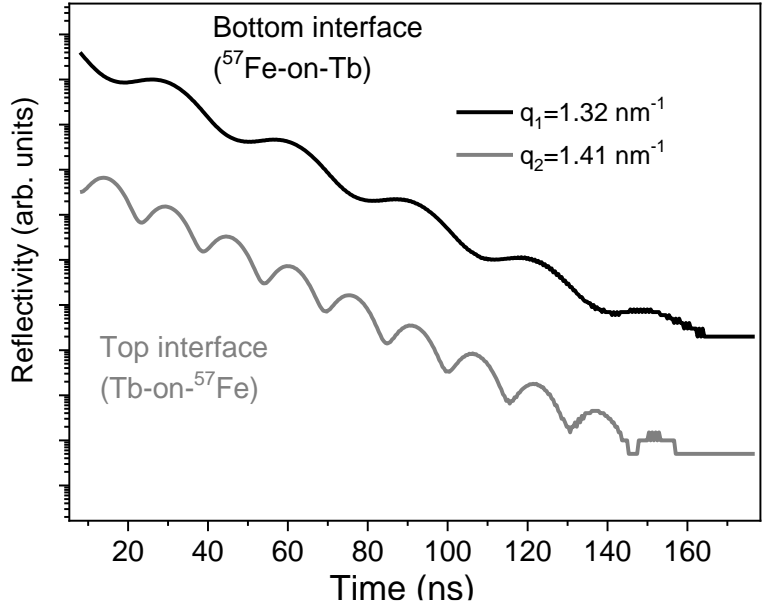

**Fig. 2** GI-NRS decay spectrum at different incident angles ($q_1$ and $q_2$) correspond to different $^{57}$Fe layers.

In conclusion, depth-resolved NRS measurements under XSW can provide information about the atomic as well as magnetic structure of both interfaces (Fe-on-M and M-on-Fe; M=Pt, Tb, etc) in the ant Fe-based trilayers. By analyzing the quantum beat pattern in the nuclear decay at different q, the hyperfine fields' magnitude and orientation at different interfaces can be accurately determined. The possibility of studying the two interfaces in a single sample (by appropriately tuning the angle of incidence) can allow researchers to follow the evolution of the two types of interfaces with *in-situ* thermal annealing. Variation in PMA with interface modification using thermal annealing can also be achieved to get insight into the origin of PMA in the potential multilayer structures.